\documentclass{elsart}
\usepackage{natbib}
\usepackage{graphicx}
\begin{document} 
\begin{frontmatter}

\title{	An algorithm for calculating the Lorentz angle in silicon detectors}

\author[iekp]{V. Bartsch}
\author[iekp]{W. de Boer}
\author[iekp]{J. Bol}
\author[iekp]{A. Dierlamm}
\author[iekp]{E. Grigoriev}
\author[iekp]{F. Hauler}
\author[iekp,cern]{S. Heising}
\author[iekp]{L. Jungermann}

\address[iekp]{Institut f\"ur Experimentelle Kernphysik, Universit\"at Karlsruhe (TH), P.O. Box 6980, 76128 Karlsruhe, Germany}
\address[cern]{European Laboratory for Particle Physics (CERN), CH-1211 Geneve 23, Switzerland}

\begin{abstract}
Future experiments will use silicon sensors in the harsh radiation environment of the LHC (Large Hadron Collider) and high magnetic fields. The drift direction of the charge carriers is affected by the Lorentz force due to the high magnetic field. Also the resulting radiation damage changes the properties of the drift. 

In this paper measurements of the Lorentz angle of electrons and holes before and after irradiation are reviewed and compared with a simple  algorithm to compute the Lorentz angle.

\end{abstract}


\end{frontmatter}

\section{Introduction}
\begin{figure}
\begin{center}
\includegraphics [width=14cm,clip]{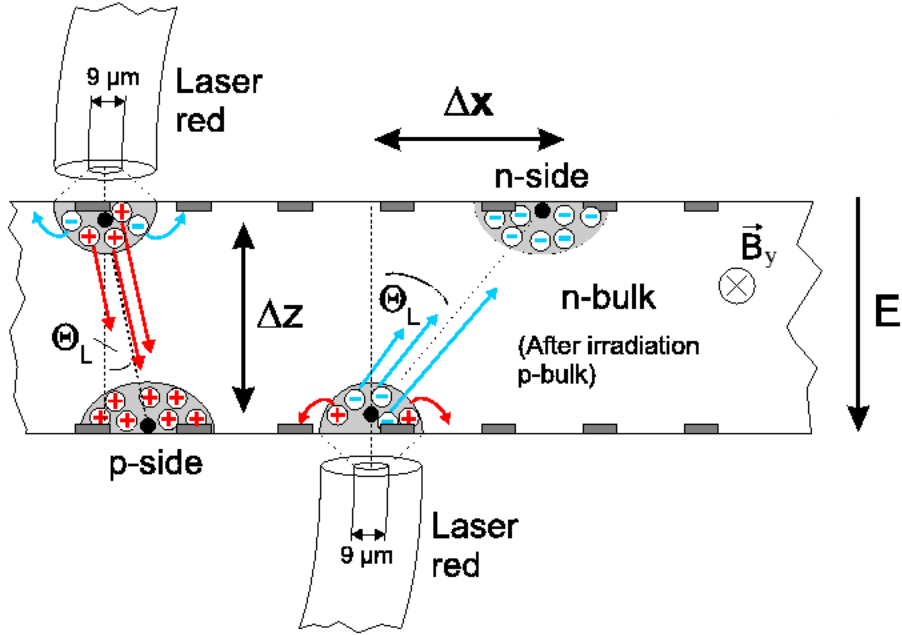}

\caption{\label{f1} \it The figure shows the principle of the Karlsruhe setup to measure the Lorentz angle of holes and electrons \cite{osaka:2000}. It is equipped with three fibers delivering laser light to the silicon. The red lasers have a penetration depth of a few $\mathrm{\mu}$m, so with a laser pulse on the n-side or  p-side one can measure the drift of electrons and holes, respectively. The infrared laser generates charge throughout the whole thickness of the detector.}
\end{center}
\end{figure}

The Lorentz angle $\Theta_{\mathrm{L}}$, by which charge carriers are deflected
in a magnetic field perpendi\-cular to the electric field, is defined by:
\begin{equation}
\label{firstequ}
\tan (\Theta_{\mathrm{L}}) = \frac{\Delta x}{d} = \mu_{\mathrm{\mbox{\tiny H}}} B = r_{\mathrm{\mbox{\tiny H}}} \mu B,
\label{eq1}
\end{equation}
where $d$ corresponds to the drift distance along the electric field
and $\Delta x$ to the shift of the signal position (see
Fig. \ref{f1}). If the ionization is produced at the surface, the
drift distance equals the detector thickness. If the ionization is
produced homogeneously by a traversing particle, the averaged drift
distance is only half the detector thickness. The drift mobility in a magnetic field, called
the Hall mobility, is
denoted by $\mu_{\mathrm{\mbox{\tiny H}}}$, the drift mobility without magnetic field by
$\mu$. They are related by the 
 Hall scattering factor
$r_{\mathrm{\mbox{\tiny H}}}=\mu_H/\mu$.  This factor describes the mean free
time between carrier collisions, which depends on the carrier energy
\cite{smith}. The Hall scattering factor has a value of $\approx
\;$0.7 for holes and $\approx \;$1.15 for electrons at room
temperature \cite{lb}. An important question is the dependence of the Lorentz shift on the irradiation dose.
Radiation damage can change the drift properties, which will change the Lorentz shift.
Such a shift appears as a misalignment of the detector as function of radiation dose,
which itself is a function of the distance from the interaction point.

\section{Experimental results}
\label{chapter4}

A comprehensive study of Lorentz shifts of non-irradiated and irradiated sensors was performed at Karlsruhe \cite{osaka:2000,roederer:1998,heising:1999,hauler:2000}. The Lorentz angle has been measured for electrons and holes separately. A temperature range of 77\,K-300\,K was covered. The Lorentz angle is measured by injecting charges at the surface on one side and observing the drift through the sensor by measuring the position of the charge on the opposite side (see Fig. \ref{f1}). Alternative methods are described in \cite{Belau:1983,Aleppo:2000}.

Charges are generated by injecting laser light with a wavelength of $\lambda$\,$\approx$\,650\,nm, which has an absorption length of $\approx$\,3\,$\mathrm{\mu}$m at 300\,K. In this case charge carriers of one type are collected at the nearest electrode, whereas the carriers of the other type drift towards the opposite side. This allows the measurement of the Lorentz angle for electrons and holes separately by injecting laser light either on the p- or n-side. The laser from Fermions Lasertech \cite{fermions} has a maximum power of 1\,mW, which could be
adjusted by the pulse height and width of the input pulse to the laserdiode. Typically a laser pulse with a width around 1\,ns was generated.The laser pulse was sent to the sensor via a single-mode fibre with an inner diameter of a few microns. So the beam spot on the detector could be varied from a few micron onwards by changing the distance between fibre and sensor.  

For the measurements the JUMBO magnet from the Forschungszentrum Karls\-ruhe \cite{jumbo} has been used in a $B$\,$\leq$\,10\,T configuration with a warm bore of 72\,mm. The sensors are double sided microstrip detectors of approximately 2x1\,cm from the HERA-B production by Sintef \cite{Abt}. They have a strip pitch of 50\,$\mathrm{\mu}$m on the p--side and 80\,$\mathrm{\mu}$m on the n--side; the strips on opposite sides are oriented at an angle of 90$^{\circ}$ with respect to each other. The capacitively coupled read out strips are connected through a 1\,M$\Omega$ resistor to the bias ring. The read out chip for the strip detectors is the Premux128-Chip with a shaping time of 45\,ns \cite{jones}. The chip's common mode is suppressed by a double correlated sampling technique, which substracts the signal's baseline. The threshold could be adjusted but the signal to noise ratio from the laser pulse, corresponding to a few times the signal from a minimum ionizing particle (MIP), was sufficiently high, so that the thresholds were not critical.

The averaged signal position $\bar{x}$ is computed from either a fit with
the sum of two Gaussians or from the center of gravity of the pulse heights $p_i$ on neighbouring strips $x_i$, i.e. $ \bar{x}  = {\sum p_i\cdot x_i}/{\sum p_i}$. Both methods gave comparable results. The first method was used.

\begin{figure}
\begin{center}
\includegraphics [width=12.5cm,clip, angle=0] {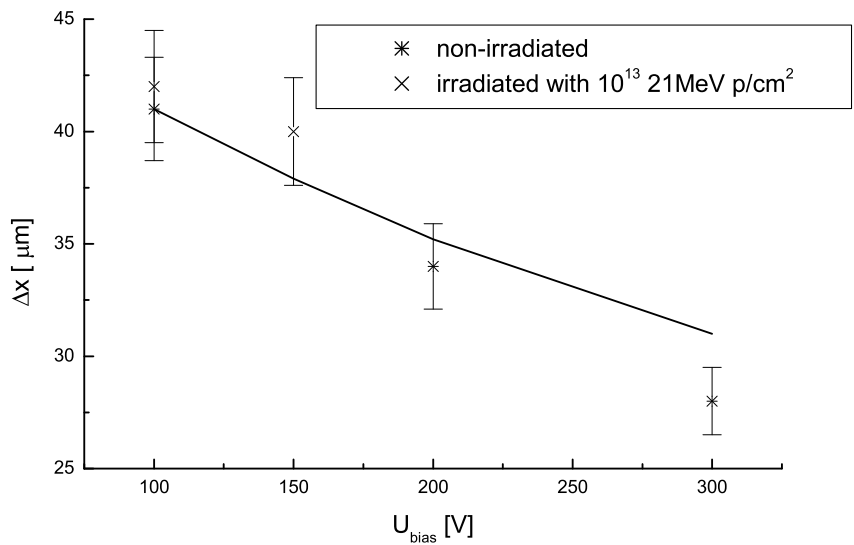}
\caption{\label{tabfig3} \it The Lorentz shift $\Delta x$ for holes generated at the surface for a 300\,$\mathrm{\mu}$m thick sensor in a 4\,T magnetic field versus bias voltage $U_{\mathrm{bias}}$. The bars represent systematic errors. The temperature is 260\,K. The line is calculated from the algorithm discussed in the text.}
\end{center}
\end{figure}

\begin{figure}
\begin{center}
\includegraphics [width=12.5cm,clip, angle=0] {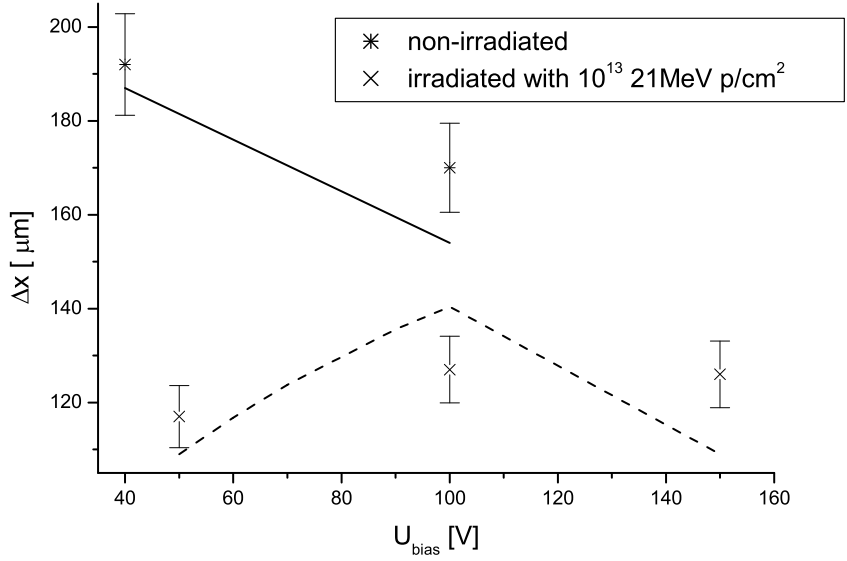}
\caption{\label{tabfig4} \it The Lorentz shift $\Delta x$ for electrons generated at the surface for a 300\,$\mathrm{\mu}$m thick sensor in a 4\,T magnetic field versus bias voltage $U_{\mathrm{bias}}$. The bars represent systematic errors. The temperature is 280\,K. The lines are calculated from the algorithm discussed in the text.}
\end{center}
\end{figure}

Our measurements showed that the Lorentz shift with the magnetic field are linear up to 9\,T \cite{osaka:2000}, which means that Eq. \ref{eq1} can be used. Before irradiation  the sensor depletes fully at a bias voltage of 40\,V, while after irradiation with 1.0$\cdot$10$^{13}$\,{21 MeV}\,protons\,/\,$\mathrm{cm}^2$ the depletion voltage  increased to 100\,V. This implies that the bulk is inverted from n-type to p-type material, as expected \cite{rd48}. The bulk damage from {21 MeV}\,protons is  about 2.1 times the damage from {1 MeV} \,neutrons \cite{huhtinen}. Numerical results on the Lorentz angles and shifts are displayed in Figs. \ref{tabfig3} and \ref{tabfig4} before and after irradiation, respectively.

The Lorentz shift for holes is not strongly depending on irradiation as shown in Fig. \ref{tabfig3}, while for electrons there is a clear dependence on the irradiation dose as shown in Fig. \ref{tabfig4}. The calculated values in the Figures are discussed in the next section. The strong decrease of the Lorentz shift for electrons after irradiation for bias voltages below 100 V originates from
the fact that the detector is not depleted anymore for these voltages, thus the Lorentz angle is only computed in the depleted zone as will be discussed in section \ref{electrons}.

\section{Algorithm for the Lorentz angle at full depletion}
\label{algorithm}

At moderate irradiation doses detectors usually can be operated at full depletion. However, for pixel detectors close to the LHC beampipe, this is not necessarily the case. We first discuss an algorithm to calculate the Lorentz angle at full depletion, while the case without full depletion will be discussed in Sect. \ref{electrons}. In this section we first discuss non-irradiated sensors and then the changes needed for moderately irradiated sensors which still can be fully depleted.

\subsection{Non-irradiated sensors}

The mobility $ \mu(E) $ is proportional to the mobility at low electric field $ \mu_{\mathrm{low}}$. It decreases with increasing electric field until it saturates. It can be parametrized as \cite{canali:1975}:

\begin{equation} \mu(E) = \frac{ \mu_{\mathrm{low}}}{(1+ ( \frac{ \mu_{\mathrm{low}}E}{v_{\mathrm{sat}} }) ^{ \beta }) ^{ \frac{1}{ \beta }}}
\label{canali}
\end{equation}
For holes the parameter values used are:
$\mu_{\mathrm{low}} = 470.5 \frac{cm^2}{Vs} \cdot \Big( {T}/{300} \Big)^{-2.5};$~ 
$\beta=1.213 \cdot \Big({T}/{300} \Big)^{0.17};$~
$ v_{\mathrm{sat}}=8.37 \cdot 10^{6} cm/s \cdot \Big({T}/{300} \Big)^{0.52},$~
while for electrons the following values are used:
$ \mu_{\mathrm{low}} = 1417 \frac{cm^2}{Vs} \cdot \Big( {T}/{300} \Big)^{-2.2};$~
$ \beta=1.109 \cdot  \Big({T}/{300}\Big)^{0.66};$~
$ v_{\mathrm{sat}}=1.07 \cdot 10^{7} cm/s \cdot \Big(\frac{T}{300}\Big)^{0.87}.$ 

The electric field depends on the z-position, the bias voltage, the full depletion voltage $U_{\mathrm{depl}}$ and the thickness of the depleted zone of the sensor d:

\begin{equation} E(z) = \frac{ U_{\mathrm{bias}} - U_{\mathrm{depl}} }{ d } + \frac{ 2 \cdot U_{\mathrm{depl}} }{ d } \cdot \Big( 1 - \frac{ z }{ d } \Big)
\label{equ2}
\end{equation}

This equation is an approximation for highly segmented detectors because of the perturbation of the field near the highly doped collection zones of a thickness of a few $\mathrm{\mu}$m. These zones are only a few percent of typical sensor thickness of 300 or 500\,$\mathrm{\mu}$m. The averaged electric field can be calculated from Eq. \ref{equ2} to be:

\begin{equation} E_{\mathrm{mean}}= \frac{E(z=0)+E(z=d)}{2}= \frac{U_{\mathrm{bias}}}{d}
\label{equ3}
\end{equation}
This E$_{\mathrm{mean}}$ is independent of the depletion voltage.
The electric field is not constant in a sensor. For full depletion it varies from $2\,\mathrm{U}_{\mathrm{depl}}/\mathrm{d}$ to zero. For large values of U$_{\mathrm{depl}}$ the electric field can change from the saturation regime to zero with a correspondingly strong change in the mobility.

\begin{figure}
\begin{center}
\includegraphics [width=13.5cm,clip, angle=0] {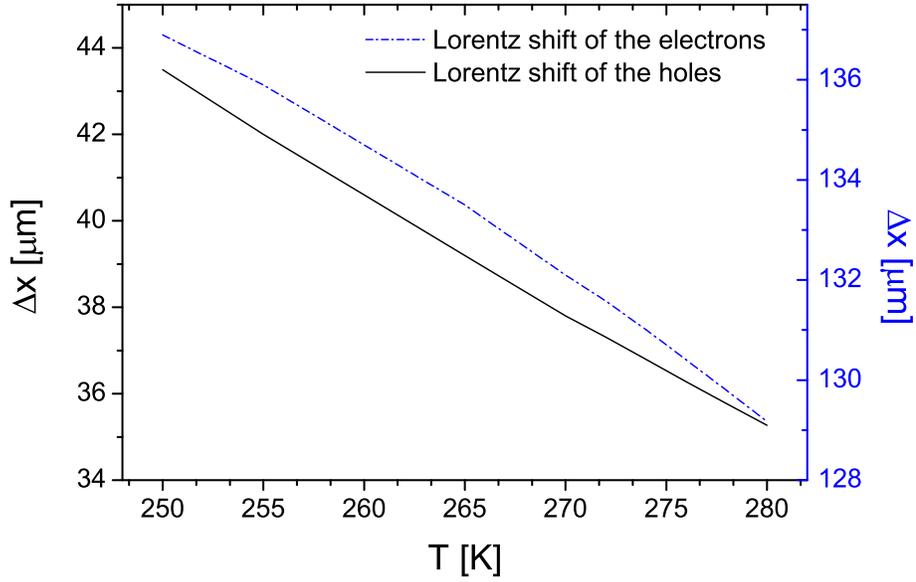}
\caption[]{\label{temperature} \it The temperature dependence of the Lorentz shift $\Delta x$ for holes (lower curve, left scale) and electrons (top curve, right scale)   predicted for a 300\,$\mathrm{\mu}$m thick sensor in a 4T magnetic field at a bias
voltage  $U_{\mathrm{bias}}$ of 100\,V.}
\end{center}
\end{figure}
\begin{figure}
\begin{center}
%\label{f6}	
\includegraphics[clip,width=6.7cm]{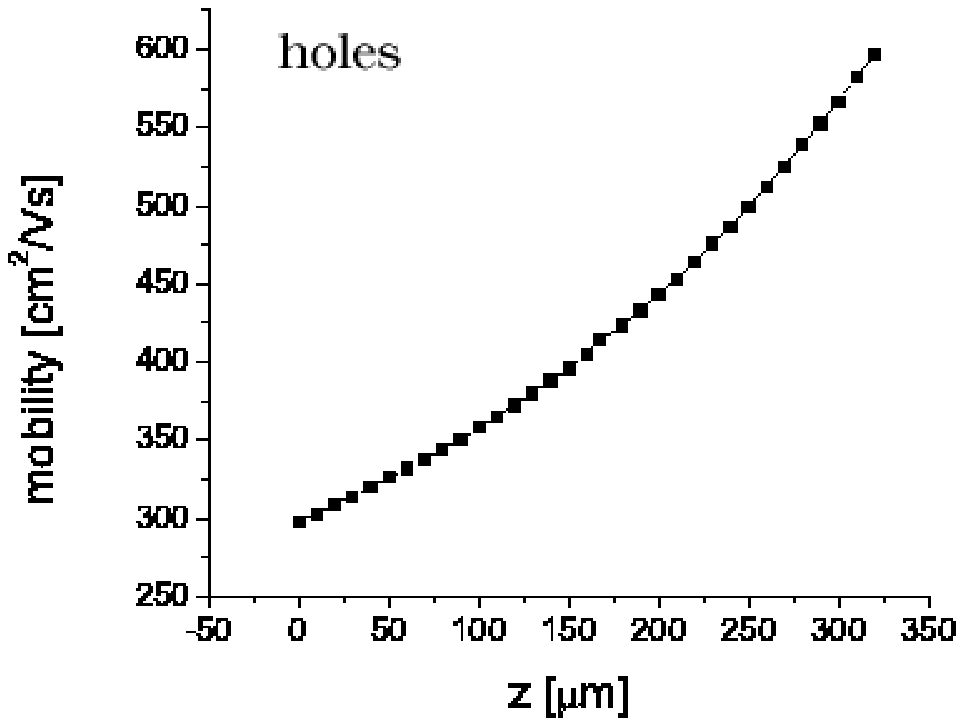}
\hspace {2mm}
\includegraphics[clip,width=6.7cm]{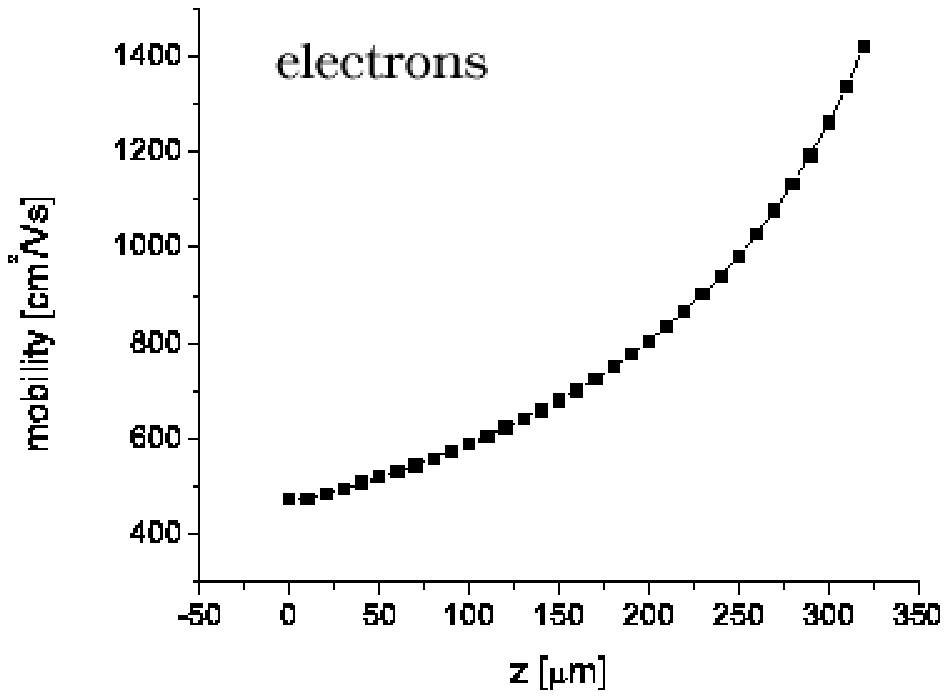}
\hspace {2mm}
\includegraphics[clip,width=6.7cm]{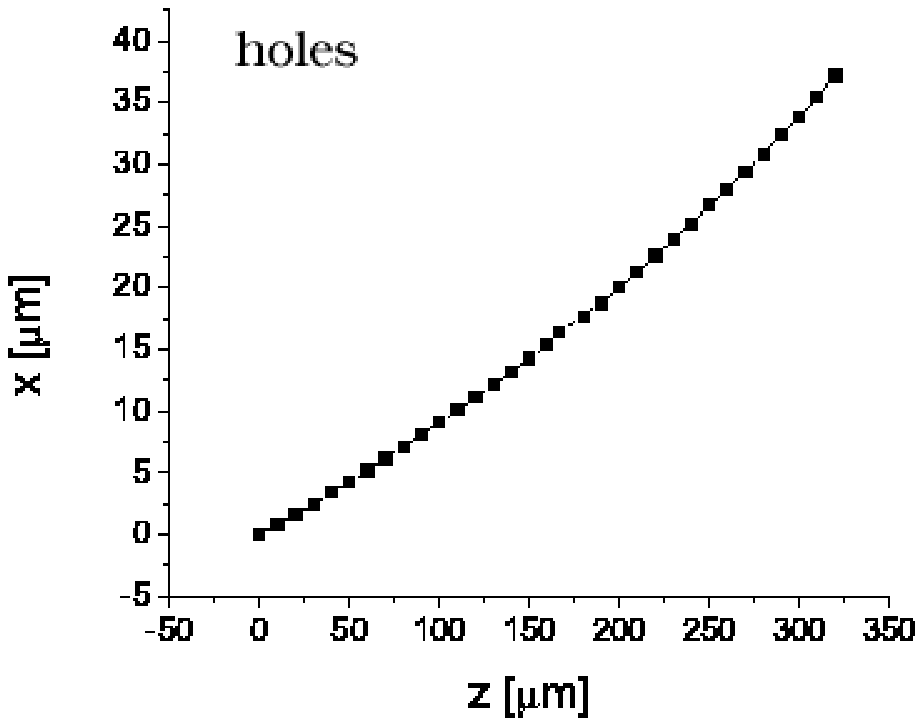}
\hspace {2mm}
\includegraphics[clip,width=6.7cm]{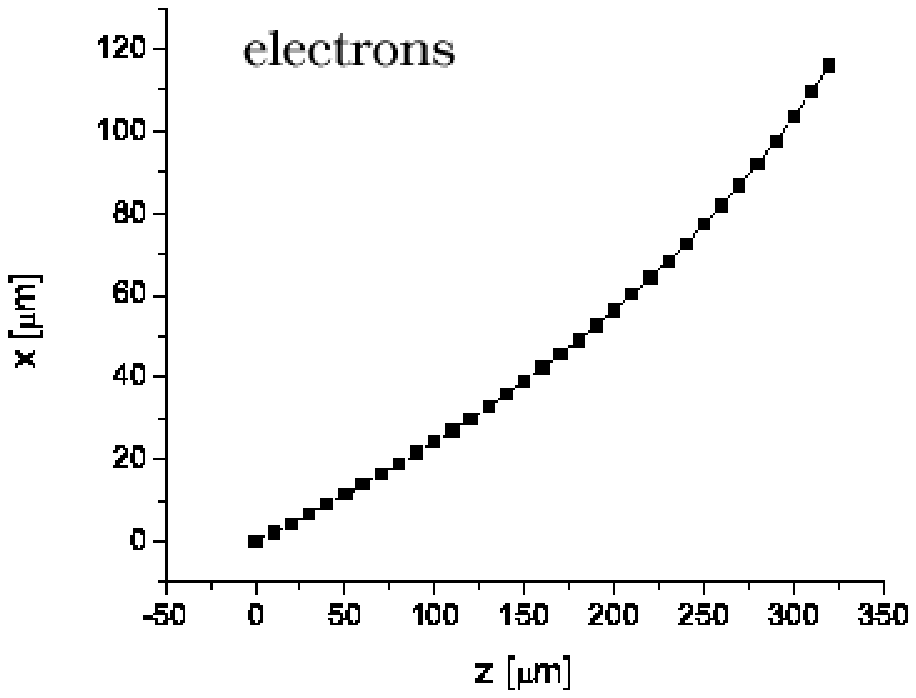}
\caption{\label{f6} \it Simulated mobility (top curve) and mean trajectory (lower figure) of holes (left hand side) and electrons (right hand side) at a full depletion voltage of 280\,V and a bias voltage of 350V at a temperature of 263\,K in a 4\,T magnetic field. The thickness is 320\,$\mathrm{\mu}$m. The total Lorentz shift for holes is 37\,$\mathrm{\mu}$m, for electrons 116\,$\mathrm{\mu}$m, the total shift calculated with a mean electric field for holes is 36\,$\mathrm{\mu}$m, for electrons 104\,$\mathrm{\mu}$m.}
\end{center}
\end{figure}

Given  the electric field, one can calculate  the mobility from Eq. \ref{canali}..
The total Lorentz shift can be otained by integrating Eq. \ref{eq1} over the thickness of the sensor using at each position the mobility given by equation \ref{canali}. For small depletion voltages the use of an averaged electric field and corresponding mobility is enough to calculate the Lorentz shift from Eq. \ref{eq1}.
Figs. \ref{tabfig3} and \ref{tabfig4} show that the calculations match the measurements quite well. The figures also show that the Lorentz shifts
vary practically linear with the bias voltage for fully depleted sensors.

The temperature dependence of the Lorentz shift  is shown in Fig. \ref{temperature}. The Lorentz shift   varies about 8\,$\mathrm{\mu}$m for a 300\,$\mathrm{\mu}$m thick sensor at 100\,V in a 4\,T magnetic field, if the temperature is varied between 250\,K and 280\,K, for both electrons and holes.

\subsection{Irradiated sensors}
\label{irradiated sensors}
There are three main parameters, which change after irradiation: the mobility, the electric field and the depletion voltage.

The dependence of the mobility on the irradiation dose is still controversial. In Ref.\cite{ere:1995} no significant changes were observed in the transport properties of both electrons and holes up to 0.5$\cdot$10$^{14}\,$\,{1 MeV}\,n/cm$^2$ and a prediction is made that a fluence of at least about 10$^{15}$\,{1 MeV}\,n/cm$^2$ is necessary to affect carrier drift mobilities significantly. In Ref. \cite{brod:1999} the mobility for both carrier types in irradiated sensors agree with those for the non-irradiated sensor within errors for fluences up to 2$\cdot$10$^{14}$\,{1 MeV}\,n/cm$^2$. In contrast a change of mobility after irradiation for holes and electrons was observed in Ref. \cite{leroy:1999}:
\begin{itemize}
\item
The mobility $\mu_{\mathrm{low}}$ of holes changes slightly from 470\,cm$^2$/Vs to about 460\,cm$^2$/Vs after irradiation to a fluence of 10$^{13}$\,{1 MeV}\,n/cm$^{2}$. Above this fluence the mobility does not change any more. The change of mobility corresponds to a change in the Lorentz shift of a few percent.\\
\item
The mobility $\mu_{\mathrm{low}}$ of the electrons changes more significantly from 1417\,cm$^2$/Vs to 1000\,cm$^2$/Vs. The reduction of the mobility is so strong that one has to take this into account when computing the Lorentz shift.
\end{itemize}

In addition, the Hall scattering factor $r_{\mathrm{\mbox{\tiny H}}}$ may change after irradiation. This is equivalent to a change in drift velocity, for what concerns the Lorentz shift. Therefore we keep $r_{\mathrm{\mbox{\tiny H}}}$ constant and fit our data with a variable mobility, which can describe the data well, as shown in Table \ref{tablowmu}. The needed shift in mobility is within the range given in Ref. \cite{leroy:1999}.
Here we used Eq. \ref{equ3} to calculate the electric field, since the depletion voltages were reasonably low. For higher depletion voltages the bias voltage is so high, that the saturation of the drift velocity has to be taken into account, which implies the use to Eq. \ref{equ2} instead of Eq. \ref{equ3}. In this case the depletion voltage has to be known.
For electrons the nonlinear saturation regime is reached at lower fields than for holes, because of the higher ratio $\mathrm{\mu}_{\mathrm{low}}/\mathrm{v}_{\mathrm{sat} }$ in Eq. \ref{canali}. This difference in nonlinearity between electrons and holes is displayed in Fig. \ref{f6} for a sensor with a large depletion voltage of 280\,V.
\begin{table}[thb]
\caption[]{\label{tablowmu} \it The Lorentz shift $\Delta x$ for electrons generated at the surface for a 300\,$\mathrm{\mu}$m thick sensor in a 4\,T magnetic field at 280\,K as function of  bias voltage. The sensor was irradiated with 21 MeV protons up to a fluence of 10$^{13}$\,cm$^2$. The full depletion voltage is 100\,V. $\Theta_{\mathrm{sim}}$ with the reduced mobility fits the data $\Theta_{\mathrm{meas}} $ better.
}\vspace*{2mm}
\begin{center}
\begin{tabular}{|c|c|c|c|c|c|c|c|}
\hline
$U_{\mathrm{bias}}$  & $\Delta x$  & $\Theta_{\mathrm{meas}} $ & $\Theta_{\mathrm{sim}} $ in $^{\circ}$ & $\Theta_{\mathrm{sim}}$ in $^{\circ}$\\
in V & in $\mathrm{\mu}$m& in $^{\circ}$ & $\mu_{\mathrm{low}}$ = 1417\,cm$^2$/Vs & $\mu_{\mathrm{low}}$ = 1100\,cm$^2$/Vs \\ 
           & &  & at 300\,K & at 300\,K \\  
\hline
\hline
 50 & 117\,$\pm$\,7 & 21\,$\pm$\,1 & 24 & 20 \\
100 & 127\,$\pm$\,7 & 23\,$\pm$\,1 & 29 & 25 \\
150 & 126\,$\pm$\,7 & 23\,$\pm$\,1 & 25 & 22 \\
\hline
\end{tabular}
\end{center}
\end{table}

In our model we also neglected the effect of double junctions after irradiation and the effect of defects (\cite{beattie:1998}-\cite{Eremin:2002}). The justification lies only in the  agreement between the model and available data
with still rather limited precision.

\section{Algorithm for the Lorentz angle at high irradiation doses}
\label{electrons}
\begin{figure}
\begin{center}
\includegraphics [width=10cm,clip] {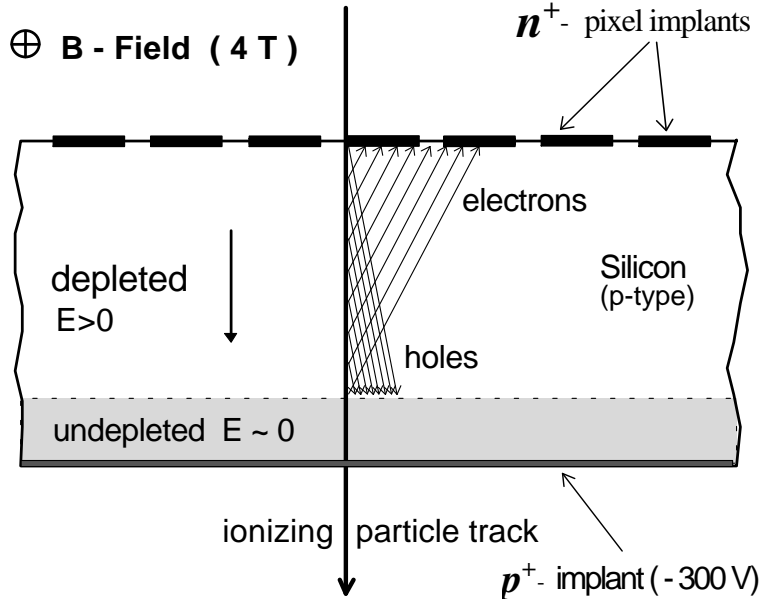}
\caption{\label{lorentzpartdep} \it After type inversion the sensor depletes from the n-pixel side. With increasing radiation dose the sensor cannot be fully depleted and the total Lorentz shift is reduced. \cite{TDR} }
\end{center}
\end{figure}
Pixel detectors are located near the beam pipe and therefore  have to stand higher radiation doses than strip detectors. It is not guaranteed that they can be always fully depleted. Since the bulk inverts after strong irradiation from n-type to p-type, the pn-junction will be on the n-side and the depletion starts from the n-side. If not fully depleted, there will be a region without electric field, in which only diffusion takes place, as indicated in Fig. \ref{lorentzpartdep}. This region will not contribute significantly to the signal, so one can consider this sensor as having a sensitive thickness corresponding to the depleted zone. We assume in our algorithm that the Lorentz shift is determined by the sensitive thickness, thus ignoring the undepleted zone. Our algorithm is compared with measurements for the ATLAS pixel sensor reported in Ref. \cite{Aleppo:2000}. The charge carriers were generated with a pion beam of 180 GeV/c. The Lorentz angle was determined by measuring the minimum mean cluster size as a function of angle of the incident beam particles. The ATLAS pixel sensors are 280\,$\mathrm{\mu}$m thick and have a full depletion voltage before irradiation of about 150\,V. The magnetic field during the measurements was 1.4\,T and
the temperature  300\,K for the non-irradiated sensors and 263\,K for the irradiated sensors. The measured values are compared to the simulated values in Table \ref{tableATLAS}. 

\begin{table}[ht]
\caption[]{\label{tableATLAS} \it The Lorentz angle for electrons
for a 280\,$\mathrm{\mu}$m thick sensor in a 1.4\,T magnetic field. The full depletion voltage before irradiation is 150\,V. The measured data are taken from \cite{Aleppo:2000}. The algorithm used for the simulation integrates the Lorentz shifts over the sensor.
The averaged Lorentz angle is defined to be the arc tangent of the total Lorentz shift divided by the depletion depth. One observes that $\Theta_{\mathrm{sim}}$ with the reduced mobility $\mu_{low}=1100 cm^2/Vs$ fits the data $\Theta_{\mathrm{meas}}$ better than the usual
value of $\mu_{low}=1417 cm^2/Vs$.
}\vspace*{2mm}
\begin{tabular}{|c|c|c|c|c|c|c|c|c|}
\hline
Fluence & $U_{\mathrm{bias}}$  & Depl. depth & $\Theta_{\mathrm{meas}} $ & $\Theta_{\mathrm{sim}}$ in $^{\circ}$ & $\Theta_{\mathrm{sim}}$ in $^{\circ}$ \\
n/cm$^2$ & in V& in $\mathrm{\mu}$m & in $^{\circ}$ & $\mu_{\mathrm{low}}$ = 1417\,cm$^2$/Vs & $ \mu_{\mathrm{low}}$ = 1100\,cm$^2$/Vs\\  
          & & & & at 300\,K & at 300\,K\\  
\hline
\hline
0                 & 150 & 283\,$\pm$\,6 & 9.0\,$\pm$\,0.9 & 8.4&  \\
\hline
5$\cdot$10$^{14}$ & 150 & 123\,$\pm$\,19 & 5.9\,$\pm$\,1.3 & 6.7 & 5.7 \\
5$\cdot$10$^{14}$ & 600 & 261\,$\pm$\,8& 2.6\,$\pm$\,0.5 & 4.4 & 3.9 \\
\hline
10$^{15}$         & 600 & 189\,$\pm$\,12 & 3.1\,$\pm$\,1.0 & 3.8 & 3.5  \\
10$^{15}$         & 600 & 217\,$\pm$\,13 & 2.7\,$\pm$\,0.8 & 3.2 & 3.0 \\
\hline
\end{tabular}
\end{table}

\begin{figure}
\begin{center}
\includegraphics [width=6.7cm,clip] {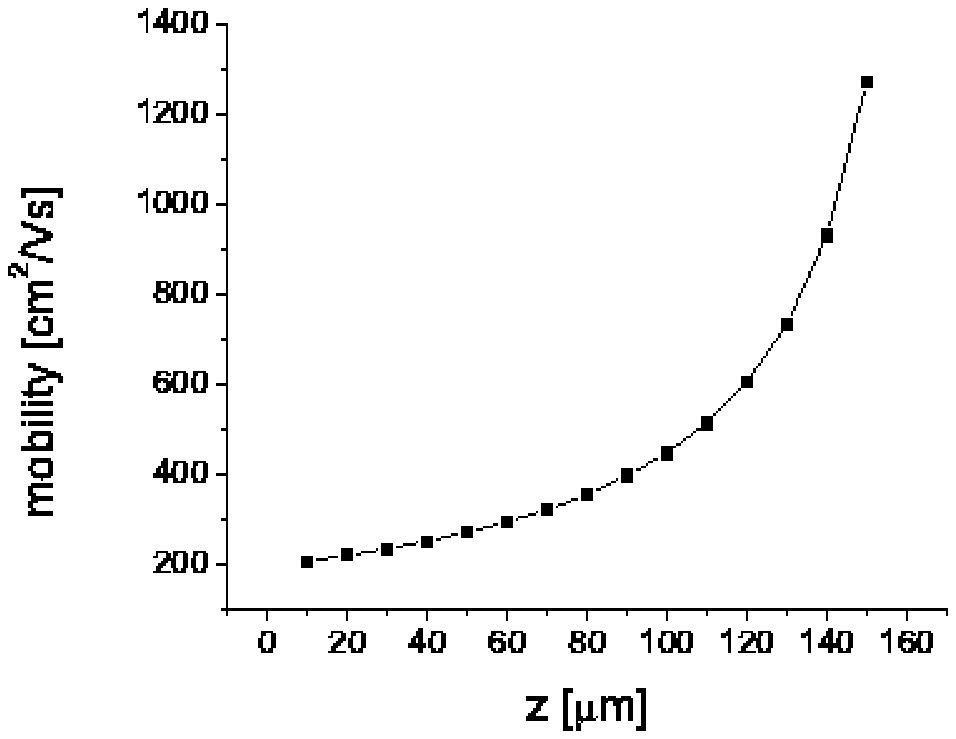}   
\hspace {2mm}
\includegraphics [width=6.7cm,clip] {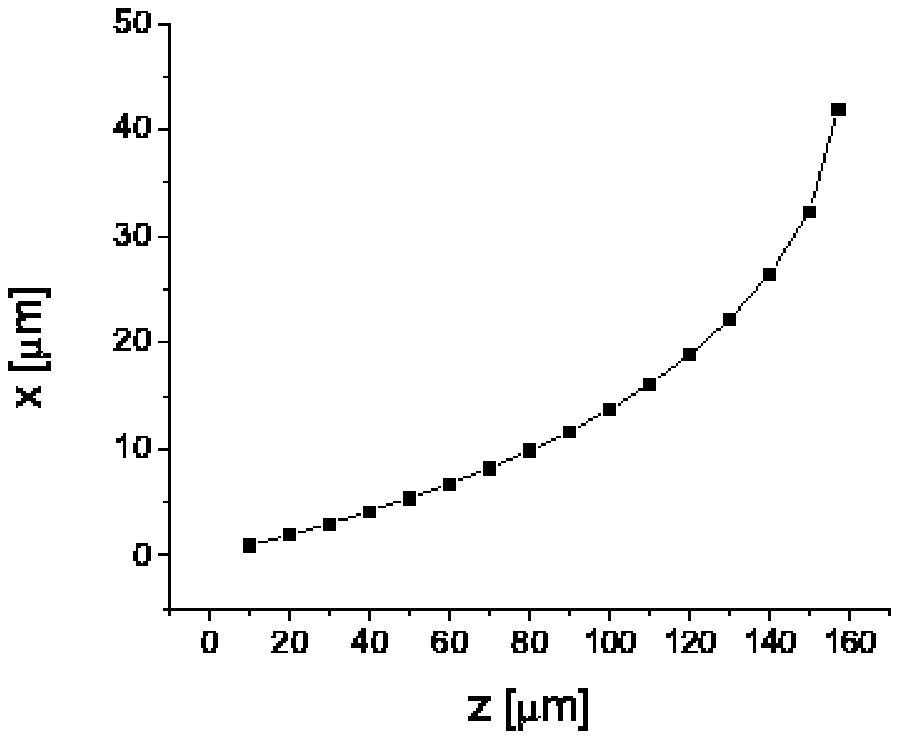}        
\caption{\label{mobel} \it Simulated mobility (left hand side) and mean trajectory (right hand side)  of electrons at a full depletion voltage of 1100\,V and a bias voltage of 300\,V at 263\,K change strongly in the depleted zone of the sensor.  }
\end{center}
\end{figure}

In the simulation the strong irradiation is taken into account by replacing the total
sensor thickness in Eq. \ref{equ2} by the depleted thickness, which is determined
by the maximum possible bias voltage. The high bias voltage leads to a strongly varying
mobility in a not fully depleted sensor,
as shown in Fig. \ref{mobel}. The trajectory of the drifting ionization
is correspondingly non-linear, so the integration of the Lorentz shifts has to be
done in sufficiently small steps. The Lorentz angle, given in Table \ref{tableATLAS},
is defined as the arc tangent of the total Lorentz shift divided by the depleted
sensor thickness.
%The electric field in the pixel sensor changes  more than in the strip sensor, if they are not fully depleted after irradiation. In that case the mobility of the electrons varies appreciably as can be seen in Fig. \ref{mobel}. Therefore the algorithm integrates over the sensor. In each part the averaged mobility and corresponding Lorentz shift was calculated using the standard and reduced mobilities $\mathrm{\mu}_{\mathrm{low}}$ of Table \ref{tablowmu} and the total Lorentz shift is the sum of the shifts in the integration parts. Throughout the sensor the Lorentz angle varies because of the changing electric field and the changing mobility (see Eq. \ref{canali}), so the Lorentz angle has to be defined as the mean Lorentz angle, calculated by the shift and the drift distance in the detector.
Comparing the simulations with the measurements one observes that the data can be fitted better with lower mobilities after irradiation, as expected from the discussion in Sect.\ref{irradiated sensors}. So it has been shown that the proposed algorithm also works for n-side
readout of not fully
depleted sensors after strong radiation, i.e.  pixel detectors,
if the reduced sensitive thickness is taken into account.

\section{Summary}

The Lorentz angle is determined by electric and magnetic fields and can be modelled by Eqs. \ref{firstequ}, \ref{canali} and \ref{equ2} or \ref{equ3}. Because of the different mobility and Hall scattering factor for holes and electrons, the Lorentz shift for electrons is at least four times the one for holes. Irradiation decreases the electron mobility  at low electric fields $\mu_{\mathrm{low}}$  significantly. The hole mobility is hardly affected by irradiation.

The simulated data have been compared with measured values from HERA-B test structures and the ATLAS pixel sensor. It has been shown that the algorithms developed here  simulate  the Lorentz shifts reasonably well, if one takes into account the reduced electron mobility
after irradiation.

\ack

This work was done within the framework of the  RD39
Collaboration \cite{rd39}.
We thank Iris Abt for supplying us with double sided mini-strip detectors.

\end{document}